\documentclass[twocolumn,aps,prd,epsf]{revtex4}
\usepackage[utf8]{inputenc}
\usepackage{epsfig}
\usepackage{amsmath}
\usepackage{dcolumn}
\usepackage[usenames]{color}
\definecolor{darkorange}{cmyk}{0,1,1,.25}

\newcommand{\be}{\begin{equation}}
\newcommand{\ee}{\end{equation}}
\newcommand{\bea}{\begin{eqnarray}}
\newcommand{\eea}{\end{eqnarray}}

\begin{document}

\title{ 
Tetraquark resonances with the triple flip-flop potential,
\\ decays in the cherry in a broken glass approximation
}

\author{P. Bicudo}
\email{bicudo@ist.utl.pt}
\author{M. Cardoso}
\email{mjdcc@cftp.ist.utl.pt}
\affiliation{Dep. Física and CFTP, Instituto Superior T\'ecnico,
Av. Rovisco Pais, 1049-001 Lisboa, Portugal}

\begin{abstract}
We develop a unitarized formalism to study tetraquarks using the triple flip-flop potential, which includes two meson-meson potentials and the tetraquark four-body potential. 
This can be related to the Jaffe-Wilczek and to the Karliner-Lipkin tetraquark models, where we also consider the possible open channels, since the four quarks and antiquarks may at any time escape to a pair of mesons.  
Here we study a simplified two-variable toy model and explore the analogy with a cherry in a glass, but a broken one where the cherry may escape from. It is quite interesting to have our system confined or compact in one variable and infinite in the other variable.
In this framework we solve the two-variable Schr\"odinger equation in configuration space.
With the finite difference method, we compute the spectrum, we search for localized states and we attempt to compute phase shifts. 
We then apply the outgoing spherical wave method to compute in detail the phase shifts and and to determine the decay widths.
We explore the model in the equal mass case, and we find narrow resonances. 
In particular the existence of two commuting angular momenta is responsible for our small decay widths. 
\end{abstract}
\maketitle

\section{Introduction}

A long standing problem of QCD is the one of the existence of localized exotic 
states and the corresponding decay to the hadron-hadron continuum. This is a
difficult problem both experimentally, since exotics usually decay to several
hadrons, and theoretically, since exotics couple to wavefunctions of at least two quarks  and two antiquarks (notice that a gluon couples to a quark and an antiquark). Nevertheless there is no QCD theorem preventing the existence of exotics, say two-gluon glueballs, hybrids, tetraquarks, pentaquarks, three-gluon glueballs, hexaquarks, etc, and the scientific community continues to search for clear exotic candidates 
\cite{:2009xt}.

In what concerns multiquarks, they may exist due to different possible mechanisms. The perspective of multiquarks as possible molecules of hadrons in attractive channels, where attraction is for instance due to quark-antiquark annihilation, has already led to the computation of decay widths, which turned out to be wide
\cite{Bicudo:2003rw,LlanesEstrada:2003us}.
Here we explore another multiquark perspective, related to the Jaffe-Wilczek model 
\cite{Jaffe:2004zg,Jaffe:2004wv,Karliner:2003dt}
where diquarks are bound directly by four-body confining potentials, produced by confining flux tubes or strings. 
In this perspective, the main effort of the scientific community has been to search for
bound states below the threshold for hadronic coupled channels, 
avoiding the computation of decay widths
\cite{Beinker:1995qe,Zouzou:1986qh}.
Aparently, the absence of a potential barrier above threshold may again produce a very large decay width to any open channel.
However Marek and Lipkin suggested that multiquarks with angular excitations may gain a centrifugal barrier,
leading to narrower decay widths
\cite{Karliner:2003dt}.

\begin{figure}[t!]
\begin{center}
\includegraphics[width=1.1\linewidth]{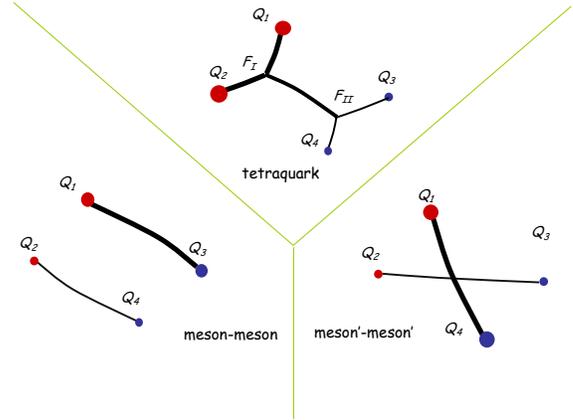}
  \caption{Triple flip-flop Potential potential. To the list of
potentials to minimize including usually only two different meson pair potentials, we join another potential, the tetraquark potential. 
\label{tripleflipflop}
}
\end{center}
\end{figure}

Here we explore the possible existence of localized tetraquarks,
or of resonances, co-existing with the continuum of open meson-meson channels. 
To address this problem, we utilize the triple flip-flop potential.
The flip-flop potential for the meson-meson interaction was developed
\cite{Oka:1985vg,Oka:1984yx,Miyazawa:1980ft,Miyazawa:1979vx,Karliner:2003dt},
to solve the problem of the Van der Waals forces produced by the two-body
confining potentials
\cite{Fishbane:1977ay,Appelquist:1978rt,Willey:1978fm,Matsuyama:1978hf,Gavela:1979zu,Feinberg:1983zz}.
Confining two-body potentials with the SU(3) colour Casimir invariant 
$ \vec \lambda_i \cdot \vec \lambda_j $ suggested by the One-Gluon-Exchange potential, 
lead to a Van der Waals potential proportional to $ V'(r) / r$ times a polarization tensor. This would lead to an extremely large Van der Waals force between mesons, which clearly does not exist. Thus two-body confinement 
dominance is ruled out for multiquark systems.
Traditionally, the flip-flop potential considers that the potential is the one
that minimizes the energy of the possible two different meson-meson configurations. This solves the problem of the Van der Waals force.
Here we upgrade the flip-flop potential, considering a third possible
configuration, the tetraquark one,
where the four constituents are linked
by a connected string
\cite{Vijande:2007ix,Vijande:2009xx}. 
The three configurations differ in the strings linking the quarks and antiquarks, 
this is illustrated in 
Fig. \ref{tripleflipflop}. 
When the diquarks $qq$ and $\bar q \bar q$ have small distances, the tetraquark configuration minimizes the string energy. When the quark-antiquark pairs $q \bar q$ and $q \bar q$ have small distances, the meson-meson configuration minimizes the string energy.
Whereas tetraquark binding 
\cite{Vijande:2007ix,Vijande:2009xx}
has been investigated with the triple flip-flop 
potential, here we address the resonance decay width whith the triple flip-flop 
potential, in particular we study localized states existing above the threshold 
of meson-meson channels opened for decay.

For the $ q_1 q_2 \bar q_3 \bar q_4$ system, there is evidence in Lattice QCD
\cite{Okiharu:2004ve},
at least for static quarks, that the hamiltonian, is
\be
	H = \sum_{i=1}^4 T_i + V_{1234}
\label{hamiltonian44q}
\ee
where the potential is decomposed in,
\be
V_{1234}= C + V_{short}+V_{conf}
\ee
in a constant term $C$, a short range screened two-body part, say 
a one gluon exchange including,
\be
V_{short} =  \sum_{i<j} \lambda_i \cdot \lambda \frac{\alpha_s}{r_{ij}}
\ee
and in a long range confining part,
\be
V_{conf} = 
	+ \sigma L_{min}( \mathbf{r}_1, \mathbf{r}_2, \mathbf{r}_3, \mathbf{r}_4 )
\ee
where $\sigma$ is the string tension and $L_{min}$ is the string lenght minimizing the energy of the colour singlet
$ q_1 q_2 \bar q_3 \bar q_4$ system. $L_{min}$ is the minimum of the three possible string lengths, 
the tetraquark string lenght
\be
L_{T_{1234}} =  r_{1\, I} + r_{2\, I}+r_{3\, II} + r_{4\, II}+r_{I\, II} \ ,
\ee
but also a meson-meson string length,
\be
L_{M_{13} M_{24}} = r_{1 \, 3} + r_{2\, 4}\ ,
\ee
and the other meson-meson string length,
\be
L_{M_{14} M_{23}} = r_{1\, 4} + r_{2\, 3} \ ,
\ee
where $r_{i \, j}$ is the distance between the points $i$ and $j$ and $I$ and $II$
are the two Fermat-Torricelli-Steiner points of the tetraquark.

\begin{figure}[t!]
\begin{center}
\includegraphics[width=1.1\linewidth]{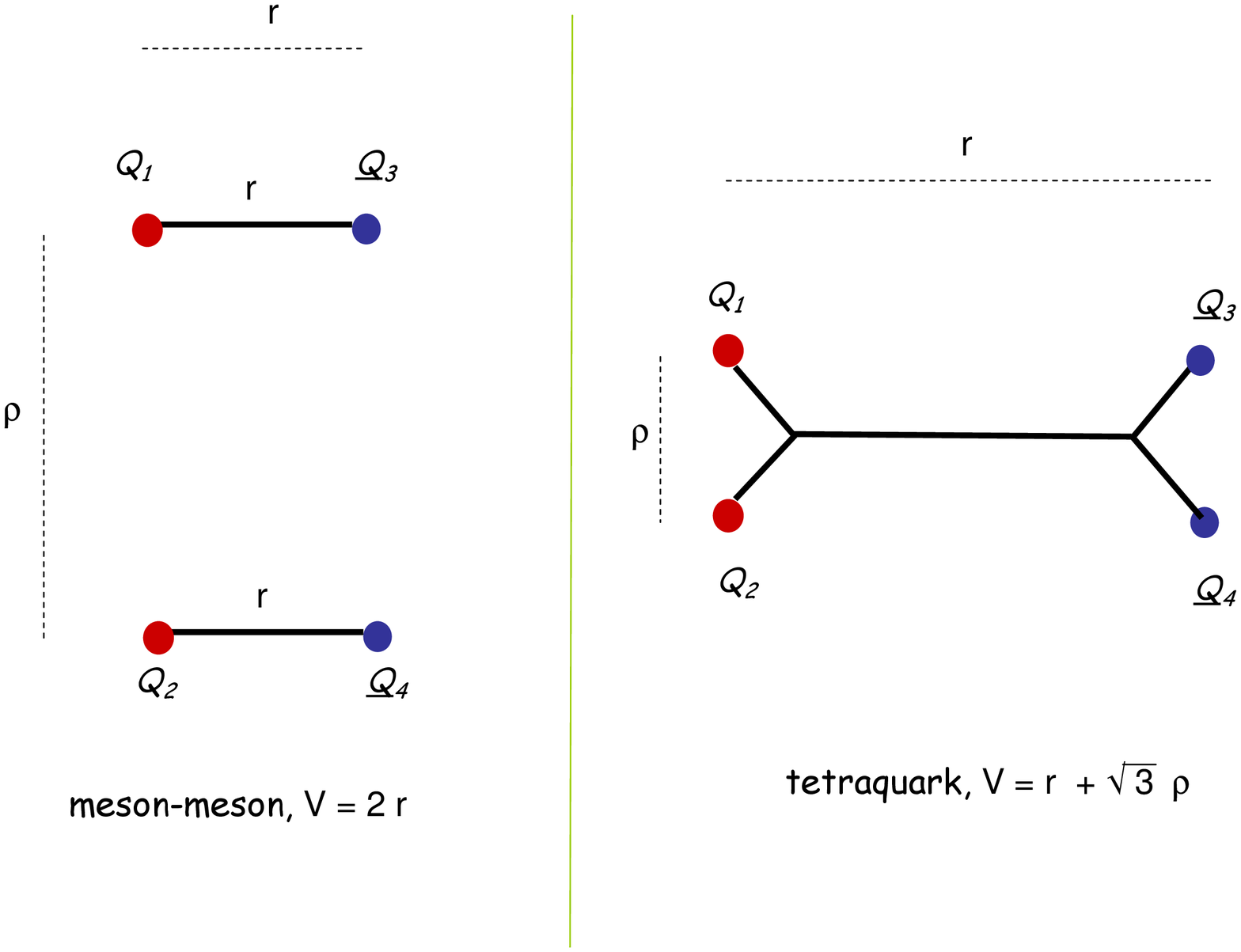}
 \caption{Simplified Potential. Assuming that the two internal mesonic
 coordinates are identical, we only have two three-dimensional variables.
 Nevertheless we remain with one confined or compact variable and one
 open variable.
\label{simplifiedpotential}
}
\end{center}
\end{figure}

However the hamiltonian of Eq. (\ref{hamiltonian44q}) is technically quite
hard to solve. Moreover we would like to capture the essence of a potential
where some dimensions are compact and others are open. Thus
for simplicity we dedicate this work to the case where,
\begin{itemize}
\item 
all the quarks and antiquarks have the same mass, although we assume that they
are different particles to avoid having to take account of exchange effects,
\item
the short range interaction and the constant term are neglected in order to single out the
effects of the confinement,
\item
the quarks are non-relativistic for simplicity,
\item 
and we also simplify the number of variables, constructing
a toy model, in order to identify the physical mechanism possibly
leading to localized tetraquark states.
\end{itemize}

In Section II we detail our toy model, analogous to a quantum cherry in a broken glass,
since our system is open to the continuum in some direction, and is confined in the other
direction. In Section III, we solve the corresponding Schr\"odinger equation, searching
for localized states although our system has no potential barrier and is open to the continuum. In Section IV we estimate the decay width of the tetraquark. In Section V we conclude.

\section{A toy model for the potential, similar to the one of a cherry in a broken glass}

\begin{figure}[t!]
\begin{center}
  \includegraphics[width=1.1\linewidth]{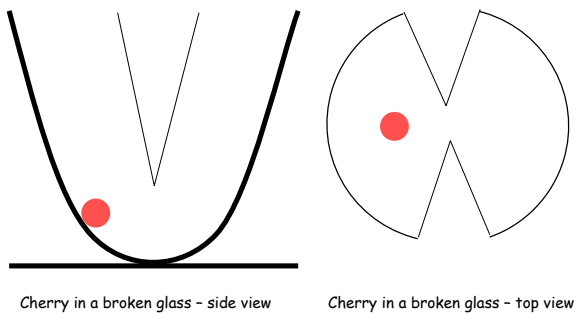}
  \caption{Cherry in a broken glass. Our simplified two-variable toy model is analogous 
to the classical mechanics textbook problem of a cherry in a glass, 
but a broken one where the cherry may escape from. 
Here we solve this model in quantum mechanics, addressing the decay widths of
a system compact in one variable and open in the other.
\label{cherrybrokenglass}
}
\end{center}
\end{figure}

Classically, a sliding cherry in a glass is equivalent, for small oscillations, to a two-dimensional harmonic
oscillator potential. However if the glass is partially broken, 
as in Fig. \ref{cherrybrokenglass}, the cherry may escape from the glass, and this is similar
to our flip-flop potential.
Note that in quantum mechanics, the text book example of a metastable system, or resonance, 
is the one of a particle decaying through a potential barrier with a tunnel effect. 
However the cherry in a broken glass is not well known in quantum mechanics, it is not clear a priori
that it leads to resonances since energetically it is completely open for the escape of the cherry
through the hole in the glass.
Nevertheless, since some classical orbits of the cherry may remain trapped in the
glass, it is interesting to study a quantum system in that class of potentials.

\begin{figure}[t!]
\begin{center}
\vspace{0.1cm}  
  \includegraphics[width=0.8\linewidth]{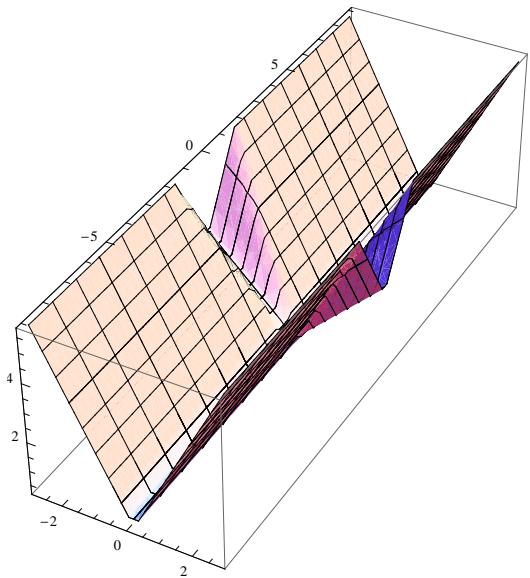}
\caption{Plot of our simplified flip-flop potential, as a function of
the two radial variables $r$ (compact) and $\rho$ (open). 
\label{flip-flop2D}
}
\end{center}
\end{figure}

We now simplify the triple flip-flop potential, in order to reduce it to a two-dimension problem, similar to the one 
of a cherry in a broken glass. It is convenient for a first study of the tetraquark
resonances, to reduce the number of variables. While a convenient set  
coordinates for the $ q_1 q_2 \bar q_3 \bar q_4$ system is,
\bea
	\boldsymbol{\rho}_{12} &=& \mathbf{r}_1 - \mathbf{r}_2 
\\
	\boldsymbol{\rho}_{34} &=& \mathbf{r}_3 - \mathbf{r}_4 
\\
	\mathbf{r}_{12,34} &=& 
{m_1 \mathbf{r}_1 + m_2 \mathbf{r}_2 \over m_1 + m_2 }
- {m_3 \mathbf{r}_3 + m_4 \mathbf{r}_4 \over m_3 + m_4 } 
\\
	\mathbf{R} &=& 
{ m_1 \mathbf{r}_1 + m_2 \mathbf{r}_2 + m_3 \mathbf{r}_3 + m_4 \mathbf{r}_4  \over  m_1 + m_2 + m_3 + m_4 }  \ .
\eea
for simplicity we assume in our simplified toy model 
that the diquark intradistance and the anti-diquark 
intradistance are similar, thus we assume that
$\boldsymbol{\rho}_{12} = \boldsymbol{\rho}_{34}$. Then we are
left with only two distances, the diquark-diantiquark distance $\mathbf r$
and the intermeson distance $\boldsymbol \rho$. Moreover, if we consider
that on average $\boldsymbol \rho$ is perpendicular to $\mathbf r$,
and considering the effect of the Fermat angle of $ 2 \pi / 3 $ in the tetraquark, 
our flip-flop potential then simplifies to two different cases, the meson-meson case
and the tetraquark case, 
\bea
V_{MM} (r, \rho) & = & \sigma (2 r) \ ,
\\
V_T (r, \rho) &=& \sigma (r + \sqrt 3 \rho) \ .
\eea
The corresponding flip-flop potential in this two-dimensional variable system,
is simply,
\be
 V_{FF}(r, \rho) =\min(V_{MM} , V_T) \ ,
 \ee
and we depict it in Fig. \ref{flip-flop2D}.

In what concerns the kinetic energy, we assume that the 
momenta of the two Jacobi coordinates $\boldsymbol{\rho}_{12} $
and $\boldsymbol{\rho}_{34} $ are identical. Thus we
finally get the Schr\"odinger equation,
\be
\left[ 
- {\hbar^2 \over 2 m} 
\left( \nabla_r^2 + \nabla_\rho^2 \right)
+ 
V_{FF}(r, \rho) 
\right] 
\psi(r, \rho) = E \psi(r , \rho)
\label{schro}
\ee
and the unitary solutions of this equation constitute the main point of our paper.

\section{Studying the spectrum of the simplified hamiltonian in a closed box with the finite difference method }

In the Schr\"odinger Equation (\ref{schro}), separating the radial from the angular
coordinates, we get for the radial component,
\bea
- \nabla_r^2 \psi &=& - \frac{1}{r} { d^2 \over d r^2 } ( r \psi ) + {l_r (l_r+1) \over r^2 } \psi
\nonumber \\
- \nabla_\rho^2 \psi &=& - \frac{1}{\rho} { d^2 \over d \rho^2 } ( \rho \psi ) + {l_\rho (l_\rho+1) \over \rho^2 } \psi
\eea
and we discretize the equation with an anisotropic spacing in $r$ and $\rho$,
\bea
& r_i = i \, a  \ & , \ \  i=0, \cdots N_r -1 \ , 
\nonumber \\
& \rho_j = j \, b  \ & , \ \  j=0, \cdots N_\rho-1 \ , 
\eea
utilizing finite differences for the kinetic energies,
\bea
{ d^2 \over d r^2 } \ u(r_i,\rho_j) &\rightarrow& {1 \over a^2} \left( u_{i-1 \, j} - 2 u_{i \, j} + u_{i+1 \, j}\right)
\nonumber \\
{ d^2 \over d \rho^2 } \ u(r_i,\rho_j) &\rightarrow&{1 \over b^2} \left( u_{i \, j-1} - 2 u_{i \, j} + u_{i \, j+1}\right)
\eea
where $u_{i\,j} = r_i \rho_j \psi_{i\,j}$.
Thus the 2-dimensional Schr\"odinger equation is equivalent to
a sparse matrix eigenvalue equation of dimension 
$ N_r \, N_\rho  \times N_r \, N_\rho  $.

Notice that the boundary three-dimensional conditions for the radial equations 
are so that $u(r , \rho)$ must vanish at the origin of both
$r$ and $\rho$, for the wavefunction $\psi$ to be regular.
It is also convenient to utilize a box quantization, so 
the wavefunction also vanishes at the maximum distances $a \, n$ and $b \, m$
we reach, respectively for $r$ and $\rho$.

\begin{figure}[t!]
\begin{center}
\vspace{1cm}  
  \includegraphics[width=0.8\linewidth]{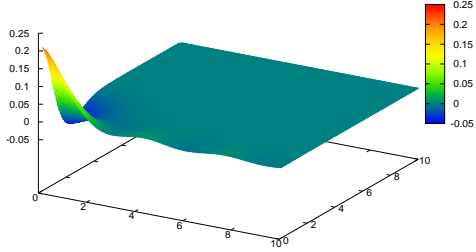}
\caption{The first localized state in the simplified 2-dimensional flip-flop potential with $C = -1.0$.
\label{firstlocalized}
}
\end{center}
\end{figure}

\begin{figure}[t!]
\begin{center}
\vspace{1cm}  
  \includegraphics[width=0.8\linewidth]{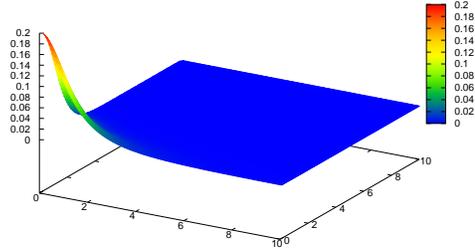}
\caption{The first bound state in the simplified 2-dimensional flip-flop potential with $C = -1.5$.
\label{boundtstateC}
}
\end{center}
\end{figure}

We solve the corresponding eigenvalue equation for a sparse matrix.
In what concerns the spectrum, we get $n \, m $ eigenstates of the hamiltonian. 
In the $\rho$ direction, to have the best possible simulation of the continuum, we use a $b \, m$ as large as 
possible. 

Most of them are continuum-like states, i.e. they extend to the end of the box, and in the limit of an infinite box their spectrum would be continuous, and only a few may be localized states.

\begin{figure}[t!]
\begin{center}
\vspace{1cm}  
  \includegraphics[width=0.8\linewidth]{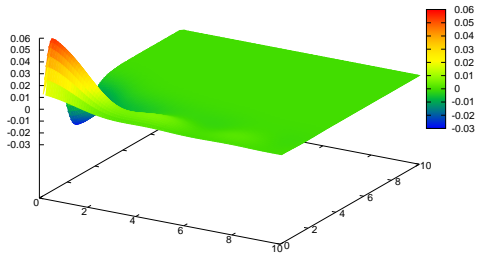}
\caption{Semi-localized state, or resonance for $l_r = 1$.
\label{localized_lr1}
}
\end{center}
\end{figure}

\begin{figure}[t!]
\begin{center}
\vspace{1cm}  
  \includegraphics[width=0.8\linewidth]{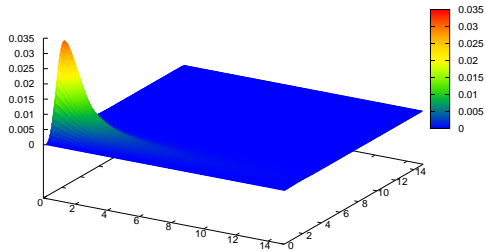}
\caption{Bound state for $l_r = 3$.
\label{boundtstate_lr3}
}
\end{center}
\end{figure}

To search for the localized solutions we compute the Radius Mean Square (RMS)  of each solution,
in the open $\rho$ direction. A localized state should have a smaller RMS than essentially
all other states. Most states are expected to be continuum states with a RMS $\simeq b \, m / 2$. 
In s-waves we find no localized solutions. But to check our formalism we can force the existence of a localized
state by adding an {\it ad hoc} negative constant $C$ to the tetraquark potential $ \sigma L_{T_{1234}}$. 
The corresponding partly localized state is depicted in Fig. \ref{firstlocalized}.
While some continuum-like tail remains in the wave-function, indicating that this is a resonance state, 
a sizeable part of it is localized close to the origin, where the tetraquark potential exists.
If we further reduce the constant $C$ we will get a confined state. In Fig. \ref{boundtstateC} we
show a bound state we obtain for the constant $C = -1.5$ .
Moreover it has been
argued by Karliner and Lipkin
\cite{Karliner:2003dt}
that higher angular waves would create a centrifugal barrier that could 
originate resonances. Indeed we find resonances and even boundstates when we consider higher $l_r$. In
Fig. \ref{localized_lr1} 
we depict a resonance occurring at $l_r=1$
and in Fig. \ref{boundtstate_lr3}
we depict a localized boundstate occurring at $l_r=3$.

To study the continuum-like solutions, it is convenient to measure their momenta, or wave number, $k$. With the momentum we can understand the spectrum and also measure phase shifts.
To measure the momentum $k$ and the phase shifts $\delta$ in each of the non decaying channels,
we have to measure the projection of the wavefunction in each one of the channels.
Note that the wavefunction can be expanded as
\be
\Psi( r, \rho ) = \sum_i \psi_i( \rho ) \phi_i( r ) \,.
\ee
The $\phi_i$ are the eigenfunctions of the hamiltonian
\be
\hat{H}_{MM} = - \frac{\hbar^2}{2 m} \nabla_r^2 + 2 \sigma r \ ,
\label{confhami}
\ee
where we use equal quark masses, $m_i = m$,
and are given in terms of the Airy function
\be
\phi_i( r ) = \mathcal{N}_i \frac{\textrm{Ai}( \frac{r}{r_0} + \xi_i )}{r} \, ,
\ee
where the $\xi_i $ are the zeroes of the Airy function.
So, the $\psi_i$ functions can be calculated by using the expression
\be
	\psi_i( \rho ) = \int d^3 r \, \phi_i(r)^* \Psi( r, \rho ) \, .
\ee
To measure the momenta $k_i$ and the phase shifts $\delta_i$, 
we simply fit the large $\rho$ region of the non-vanishing $\psi_i$ to
the expression
\be
\psi_i \rightarrow A_i \frac{\sin( k_i \rho + \delta_i )}{ \rho } \ .
\ee

\begin{figure}[t!]
\begin{center}
\vspace{0.1cm}  
  \includegraphics[width=0.8\linewidth]{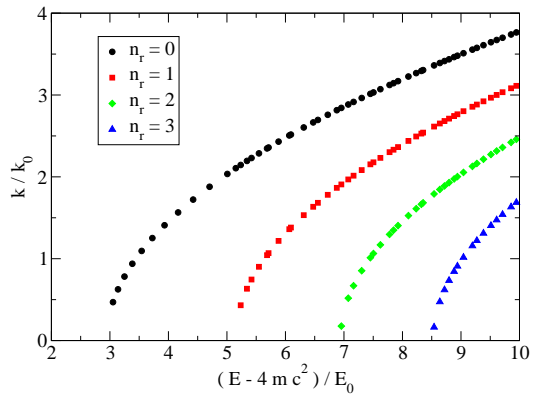}
\caption{ Momenta of the various components as a function of the energy.
\label{momenta}
}
\end{center}
\end{figure}

As can be seen in Fig. \ref{momenta}, the momenta $k_i$ obey the relation
\be
k_i(E) = \sqrt{ 2 ( E - \epsilon_i ) } ,
\ee
where $\epsilon_i$ is the threshold energy of the respective channel.
Note that there are different thresholds opening at higher momenta, 
corresponding to radial excitations in the $r$ variable.

We also show the corresponding results for the phase shifts 
$\delta_i$ are shown in Fig. \ref{phaseshifts}.
As can be seen, the behaviour is irregular when we have more than one
channel, this is due to the different contributions of multiple channels, 
for each eigenstate calculated in the finite difference scheme. 

While the finite difference method is adequate to observe the quasi-localized
resonances and the localized boundstates, and also to measure the
momenta of the continuum states, we now move on to a different method
with the aim to study in detail the phase shifts.

\section{Studying the phase shifts via the outgoing spherical wave method}

\begin{figure}[t]
\begin{center}
\vspace{1cm}  
  \includegraphics[width=0.8\linewidth]{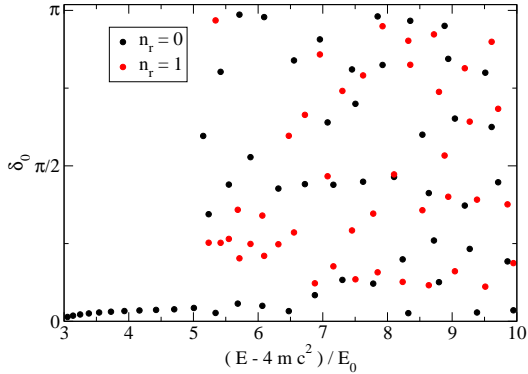}
\caption{ "Phase shifts" obtained from the finite differences ( by projecting the eigenstates
in the meson-meson eigenstates ). As can be seen the behaviour is irregular when we have more than one
channel, this is due to the different contributions of multiple channels, for each eigenstate calculated
in the finite difference scheme.
\label{phaseshifts}
}
\end{center}
\end{figure}

The irregularity of the phase shifts extracted from the finite
difference solutions in a closed box, 
show in Fig. \ref{phaseshifts}, motivates us
to utilize a precise and unitarized method to compute the phase shifts. 
We now solve the Schr\"odinger equation with spherical
outgoing waves, to explicitly compute the phase shifts.

\subsection{method}


In a coupled channel problem, the Schr\"odinger equation is,
\be
- \frac{\hbar^2}{2 m_i} \nabla^2 \Psi_{i} + V_{ij} \Psi_j = ( E - \epsilon_i ) \Psi_i \ .
\label{schro_cc}
\ee
Considering the scattering from the channel $i$ to the channel $j$, 
we have the following asymptotic relations, for the $i$ channel,
\be
\Psi_i \rightarrow e^{i k_i z} + f_{ii}( \hat{r} ) \frac{e^{ i k_i r } }{ r } \ ,
\label{autoscatter}
\ee
and, for $j \neq i$,
\be
\Psi_j \rightarrow f_{ij}( \hat{r} ) \frac{e^{ i k_j r } }{ r } \ , 
\label{otherscatter}
\ee
if the channel $j$ is open, otherwise vanishing.
With the conservation of the probability, we obtain the optical theorem,
\be
\sum_j \sigma_{i \rightarrow j} = \frac{4 \pi}{ k } \Im f_{ii}(0) \ .
\label{optic}
\ee
This can be formulated in terms of partial waves, in which case the Eqs.
(\ref{autoscatter}) and (\ref{otherscatter}) become,
\be
u_i^l \rightarrow \sin ( k_i r + l \pi ) + ( 2 l + 1 ) f_{ii}^l e^{i k_i r} \ ,
\ee
and
\be
u_j^l \rightarrow ( 2 l + 1 ) f_{ij}^l e^{i k_j r} \ ,
\ee
and the optical theorem becomes
\be
\sum_j \sigma_{i \rightarrow j}^l = \frac{1}{k} ( 2 l + 1 ) f_{ii}^l \ .
\label{optic2}
\ee
The $f_{ij}^l$ can be computed by considering outgoing solutions of the Eq. (\ref{schro_cc}) ,
\be
\Psi_i = e^{i k_h r} \delta_{ih} + \chi_i \ .
\ee
We have the equation for $\chi_i$
\be
- \frac{\hbar^2}{2 m_i} \nabla^2 \chi_{i} + V_{ij} \chi_j = ( E - \epsilon_i ) \chi_i - V_{ih} e^{i k_h r} \ ,
\ee
and it  can be simplified if $V_{ij}$ is spherically symmetric,  writing $\chi_i$ as,
\be
\chi_i = \frac{u_{l}(r)}{r} Y_{l m}(\theta,\varphi) \ ,
\ee
in which case the equation becomes
\be
- \frac{\hbar^2}{2 m_i} \frac{d^2 u_l}{dr^2} + V_{ij} u = ( E - \epsilon_i ) u_l - V_{ih} j_l( k r ) r \ .
\label{theeq}
\ee
Thus, to calculate the phase shifts and the cross sections we only need to solve Eq. (\ref{theeq})
equation and to analyze
the assimptotic behaviour of each component of the wavefunction.

We can reduce our problem in the dimensions 
$\rho, \, r$ to a one-dimensional problem in $\rho$ but with of coupled channels. 
We just have to expand the two-dimensional wavefunction as
\be
\Psi( \mathbf{r}, \boldsymbol{\rho} ) = \sum_i \psi_i( \boldsymbol{\rho} ) \phi_i( \mathbf{r} ) \ ,
\ee
where the $\phi_i$ are the eigenfunctions of the $r$ confined hamiltonian
of Eq. (\ref{confhami}).
The one-dimensional potentials $V_{ij}$ are given by
\be
V_{ij}( \rho ) = \int d^3 \mathbf{r} \quad \phi_i^* ( r ) ( V_{FF}( r, \rho ) - V_{MM}(r) )\phi_j ( r )
\ee
where we subtract $\hat{H}_{MM}$ from the hamiltonian, since 
$\hat{H}_{MM}$ is already accounted for in its eigenvalues $\epsilon_i$,
to be used in Eq. (\ref{schro_cc}).

Importantly, note that we have two distinct angular momenta, which are both conserved, $\mathbf{L}_r = \mathbf{r} \times \mathbf{p}_r$
and $\mathbf{L}_\rho = \boldsymbol{\rho} \times \mathbf{p}_\rho$.
So, each assimptotic state is indexed by its angular momentum $l_r$ and its radial number $n_r$, and the scattering
partial waves are indexed by $l_\rho$.
Thus the system can be diagonalized not only in the scattering angular momenta $\mathbf{L}_\rho$ but also on the confined
angular momenta $\mathbf{L}_r$.
We can describe the scattering process with four quantum numbers: The scattering angular momentum $l_\rho$, the confined
angular momentum $l_r$ and the initial and final states radial number in 
the confined coordinate $r$, $n_i$ and $n_j$.

\subsection{phase shifts}


We now compute the phase shifts, in order to search for resonances in our simplified flip-flop model.
Solving Eq. (\ref{theeq}) for this system we can compute the partial cross sections and the total cross section
for the partial wave $l$ --- either directly or by using the optical theorem of Eq. (\ref{optic2}) ---
and determine the phase shifts as well.

Note that our flip-flop potential has the same scales
of the simple Schr\"odinger equation for a linear potential, which becomes dimensionless when
we substitute ,
\bea
\left( m \sigma \over \hbar \right)^{1 /3} r  &\to& r \ , 
\nonumber \\
\Big( \frac{m}{\hbar^2 \sigma^2} \Big)^{1/3}  E &\to& E \ .
\eea
Thus the number of boundstates or resonances is independent both of the quark mass $m$
and of the string constant $\sigma$. Also, in the remaining figures, 
the energy will be divided by the non-relativistic energy scale
\be
E_0 = \Big( \frac{\hbar^2 \sigma^2}{m} \Big)^{1/3} \, , \label{enscale}
\ee
and thus our figures are dimensionless.
Note that $E_0$ is also the only energy scale we can construct 
with $\hbar$, $\sigma$ and $m$, the three
relevant constants in the non-relativistic region.

\subsubsection{the centrifugal barrier effect}

\begin{figure}[t!]
\begin{center}
\vspace{0.1cm}  
  \includegraphics[width=0.8\linewidth]{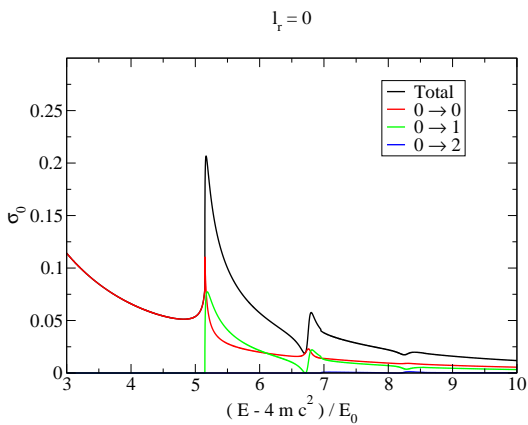}
\caption{ S-wave scattering cross sections from the channel with $l_r = 0$ and $n_r = 0$.
\label{crosssection_c0_lr0_lrho0}
}
\end{center}
\end{figure}

\begin{figure}[t!]
\begin{center}
\vspace{0.1cm}  
  \includegraphics[width=0.8\linewidth]{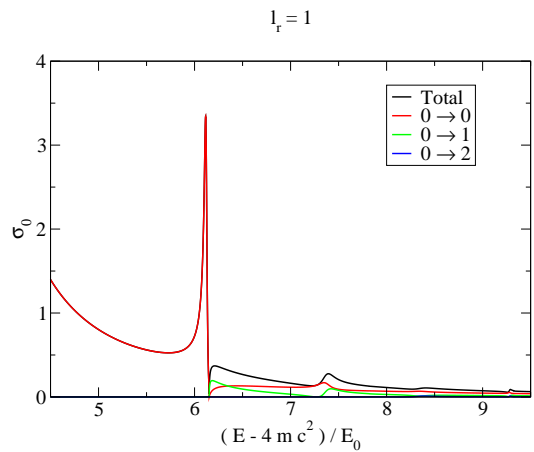}
\caption{ S-wave scattering cross sections from the channel with $l_r = 1$ and $n_r = 0$.
\label{crosssection_c0_lr1_lrho0}
}
\end{center}
\end{figure}

On Fig. \ref{crosssection_c0_lr0_lrho0} we can show the $l_\rho = 0$ partial cross sections for the scattering
from the channel with $l_r = 0$ and $n_r = 0$. In Fig. \ref{crosssection_c0_lr1_lrho0} we show the same
results but for $l_r = 1$.
Interestingly, the bumps in the cross section seem to occur prior to
the opening of a new channel.

\begin{figure}[t!]
\begin{center}
\vspace{0.1cm}  
  \includegraphics[width=0.8\linewidth]{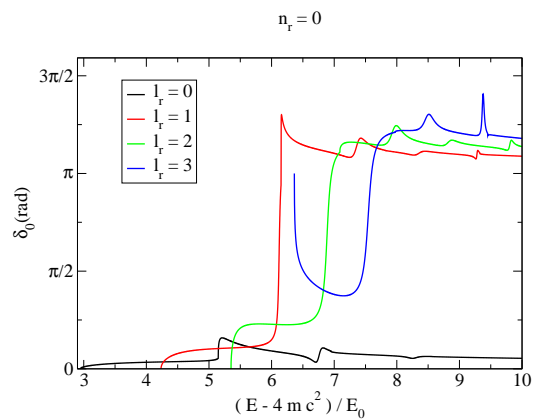}
\caption{ Comparision of the phase shifts for $l_r = 0, 1, 2$ and $3$, with $n_r = 0$.
\label{delta_c0_lrho0}
}
\end{center}
\end{figure}

\begin{figure}[t!]
\begin{center}
\vspace{0.1cm}  
  \includegraphics[width=0.8\linewidth]{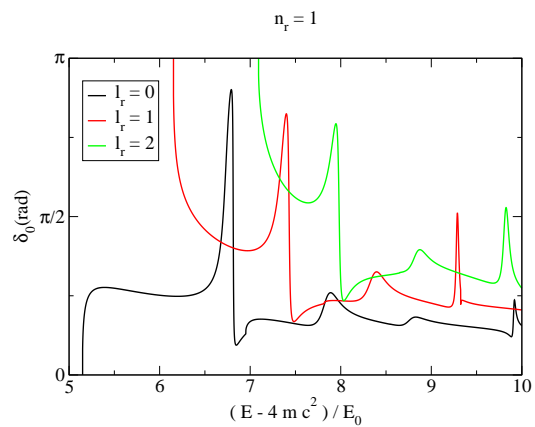}
\caption{ Comparision of the phase shifts for different $l_r$, with $n_r = 1$.
\label{delta_c1_lrho0}
}
\end{center}
\end{figure}

\begin{figure}[t!]
\begin{center}
\vspace{0.1cm}  
  \includegraphics[width=0.8\linewidth]{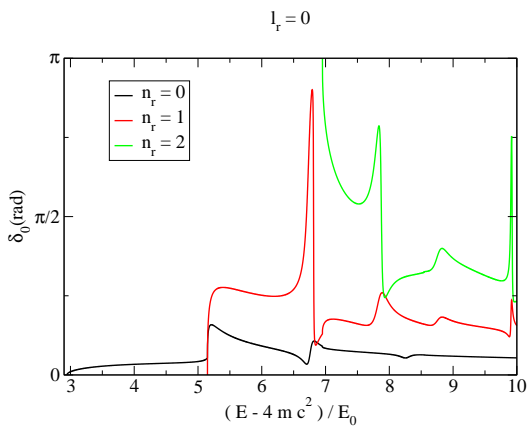}
\caption{ Comparision of the phase shifts for $l_r = 0$ and different $n_r$.
\label{delta_lr0_lrho0}
}
\end{center}
\end{figure}

In Fig.\ref{delta_c0_lrho0} we compare the phase shifts for different values of $l_r$,
namely for $l_r = 0, 1, 2$ and $3$.
For $l_r = 0$, we don't observe a resonance, since the phase shift doesn't even cross $\pi/2$.
However, for the $l_r = 1$ and $l_r = 2$ cases, the phase shifts clearly cross the $\pi/2$ line, and a resonance is formed.
This behaviour is somewhat expected, since a centrifugal barrier in $r$ would, in the case of a true tetraquark,
maintain the two diquarks separated, favouring the formation of a bound state.
The tendency of greater stability for greater orbital angular momenta seems to be further confirmed by the
$l_r = 3$, where besides the resonance, a true bound state seems to be formed, as can be seen by the
different qualitative behaviour of the phase shifts for this case.
This bound state formation confirms our observation of a localized state, in Section III, with the finite difference simulation.
The resulting wavefunction is presented in the Fig. \ref{boundtstate_lr3}.
We also show the phase-shifts for scattering in the $n_r = 1$ and various $l_r$ if Fig. \ref{delta_c1_lrho0},
and for scattering with $l_r = 1$ and various $n_r$ channels in Fig. \ref{delta_lr0_lrho0}.

\subsubsection{including an attractive constant}

As a verification of our method and of our numerical codes, 
we also study the effect of including an arbitrary constant in the tetraquark sector of the flip-flop potential.
We consider a potential of the kind
\be
V = \sigma \min( 2 r , \sqrt 3 \rho + r + C ) \, .
\ee
The use of a negative value for $C$ should result in favouring binding within the tetraquark sector and, possibly, the
appearance of a resonance, or of a bound state, in the spectrum.

\begin{figure}[t!]
\begin{center}
\vspace{0.1cm}  
  \includegraphics[width=0.8\linewidth]{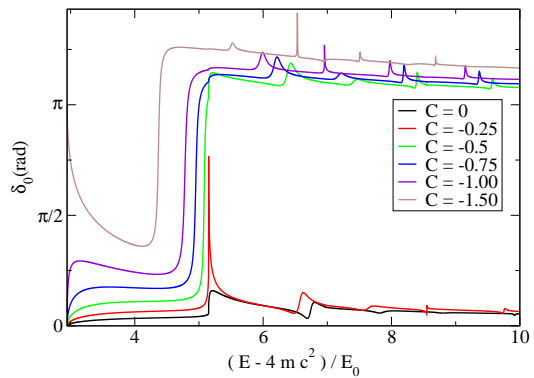}
\caption{ Effect of a different constant on the "tetraquark" sector of the potential in the
phase shift.
\label{desnivel}
}
\end{center}
\end{figure}

In Fig. \ref{desnivel} we see the results for the $l_\rho = 0$ phase shift for $l_r = 0$ and $n_r = 0$ channel.
As can be seen the phase shifts become greater for more negative values of $C$. For $C = -0.25$ we almost have the
formation of a true resonance, with $\delta$ passing $\pi / 2$ but falling shortly after. However, for $C = -0.50$ the
formation of a resonance is clear, with a jump of $\pi$ in $\delta$. 
If we further enhance the negative constant $C$ in the potential, 
we then see the appearance of a bound state.
The formation of a resonance and then of a bound state with the increase of the absolute value of $C$
was already observed in the previous section, using the finite difference approach.
The plots of the wavefunctions of the resonance and of the bound state, are shown on Figs. \ref{firstlocalized}
and \ref{boundtstateC}.
This expected result of increasing the attraction, confirms that the similar states observed while increasing the
angular momentum are indeed resonances or boundstates.

\subsubsection{a relativistic correction to the binding energy}

Another source for binding in present in relativistic corrections to the kinetic energy.
A completely non-relativistic approach leads to an error in the kinetic energy.
Note that although we have different scattering meson states, 
their non-relativistic kinetic energy is,
\be
E_i(k) = E_i^0 + \frac{k^2}{2 \mu}
\ee
while the correct non-relativistic dispersion relation for the mesons should be
\be
E_i(k) = M_i c^2 + \frac{k^2}{2 \mu_i} 
\ee
where $\mu_i$ is the reduced mass of the system, depending on the mass of the mesons 
and thus depending on their binding energy.
In the full tetraquark problem this would be a little more complex, 
since we would have more degrees of freedom, namely
$
E_{ij}(k) = ( M_i + M_j ) c^2 + \frac{k^2}{2 \mu_{ij}}
$
with
$
 \mu_{ij} = \frac{M_i M_j}{M_i + M_j} 
$ .
In our model we only need to consider the case $M_j = M_i$, since the two mesons
are no longer independent.
We introduce the relativistic correction in the mass
\be
\mu_i = \frac{M_i}{2}
\ee
where $M_i$ is given by
\be
M_i = 2 \, m + \frac{B_i}{2 c^2}
\ee
and where $B_i$  is the binding energy, given by the eigenvalues of $\hat{H}_{MM}$.
The factor  $1/2$ comes from dividing it equally by the two mesons.
So, to each state $i$ corresponds the reduced mass 
\be
\mu_i = m + \frac{B_i}{4 c^2} \, .
\ee

Without this relativistic correction, our sistem would depend 
only in three constants $\hbar$, $m$ and $\sigma$, and changing the mass would only change the
scale of the physical results, $E_0$. Now we have another constant, the speed of light, $c$,
and this we will obtain qualitatively different results for different values of $m / \sqrt{\sigma}$.

\begin{figure}[t!]
\begin{center}
\vspace{0.1cm}  
  \includegraphics[width=0.8\linewidth]{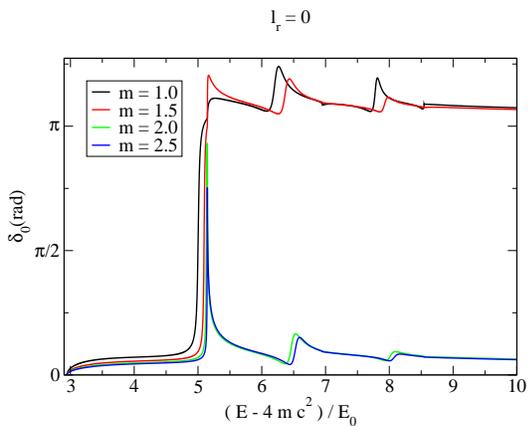}
\caption{ Phase shifts, including the reduced mass correction for $l_r = 0$.
\label{redmass_lr0}
}
\end{center}
\end{figure}

\begin{figure}[t!]
\begin{center}
\vspace{0.1cm}  
  \includegraphics[width=0.8\linewidth]{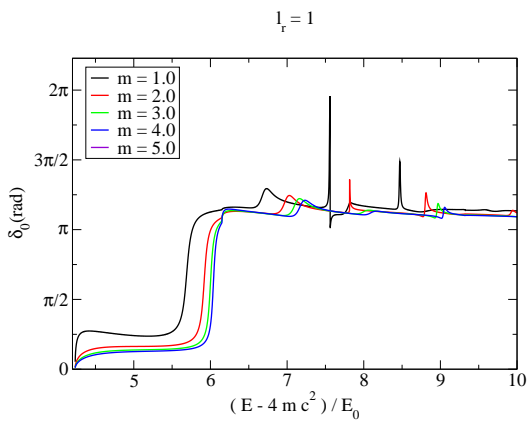}
\caption{ Phase shifts, including the reduced mass correction for $l_r = 1$.
\label{redmass_lr1}
}
\end{center}
\end{figure}

\begin{figure}[t!]
\begin{center}
\vspace{0.1cm}  
  \includegraphics[width=0.8\linewidth]{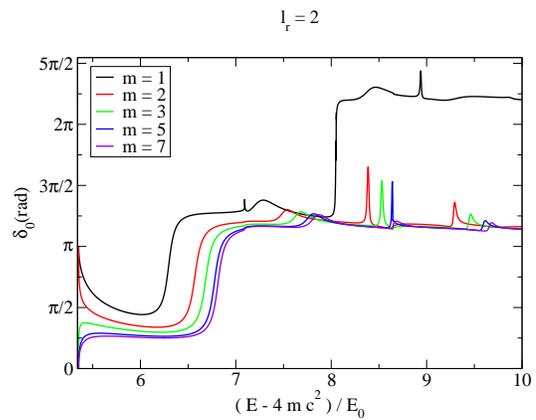}
\caption{ Phase shifts, including the reduced mass correction for $l_r = 2$.
\label{redmass_lr2}
}
\end{center}
\end{figure}

Indeed we observe that the relativistic correction enhances binding.
In Figs. \ref{redmass_lr0}, \ref{redmass_lr1} and \ref{redmass_lr2} for the $l_\rho = 0$
phase shifts of the $l_r = 1,2,3$ and $n_r = 0$ state
as a function of the scattering energy, for different values of $m / \sqrt{\sigma}$, are shown.
When $l_r = 0$ we see that for low quark masses, a resonance is present, but for a mass as low as
$2 \sqrt{\sigma}$ it is already destroyed.
For $l_r = 1$ we have a perennial resonance, which becomes thinner by increasing the quark mass.
For $l_r = 2$ if we the mass is sufficiently low, we can observe, not only one resonance, but also
a bound state and a second resonance.

\subsubsection{computing the decay widths}

Finally we can compute the decay widths of the observed resonances
utilizing the derivative of the respective phase shift,
\be
{\Gamma \over 2} = \left( d \delta \over d E \right)^{-1} \Biggl|_{\delta ={\pi \over 2}} \ .
\ee

The results without and with the reduced mass correction are given
in Tables \ref{widthtable1} and \ref{widthtable2}.
Clearly, the width becomes larger for greater angular excitations,
and for smaller quark masses.

Notice that this is a toy model, but nevertheless we can
estimate the physical scale of our decay widths.
For instance in the case of light quarks, both our physical scales are similar 
$m \simeq \sqrt{\sigma} \simeq 0.4$ GeV,
Eq. \ref{enscale} yields an energy scale of $E_0 \simeq 0.4$ GeV.
Thus we get, for $l_r = 1$, a decay width of the order
\be
\Gamma_l \simeq 0.037 \times 0.4 \mbox[GeV] \simeq 15 MeV \, .
\ee
For tetraquarks composed of charm quarks, with a mass of approximately $1.5$ GeV bottom quark, 
the energy scale is $E_0 \simeq 0.26$ GeV
\be
\Gamma_c \simeq 10 MeV \, ,
\ee
and for a tetraquarks composed of bottom quarks, with a mass of approximately $5 GeV$, the energy scale is
$E_0 \simeq 0.17$ GeV and so the width is
\be
\Gamma_b \simeq 6 MeV \, .
\ee
As for the case of $l_r = 2$, the decay widths are 
$\Gamma \simeq 0.131 \times 0.4$ GeV $ \simeq 52 $ MeV for light quarks, 
$\Gamma \simeq 34$ MeV for the charm quark,
and $\Gamma \simeq 23$ MeV for the bottom quark.
For $l_r = 3$ the decay widths $\Gamma$ are $140$ Mev, $90$ Mev and $60$ MeV respectively.

\begin{table}
\begin{tabular}{|c|c|c|}
	\hline
	$l_r$ & $(E - 4 m c^2)/E_0$ & $\Gamma$ / $E_0$ \\
	\hline
	1 & 6.116 & 0.037 \\
	2 & 6.855 & 0.131 \\
	3 & 7.462 & 0.352 \\
	\hline
\end{tabular}
\caption{Decay widths as a function of $l_r$.}
\label{widthtable1}
\end{table}

\begin{table}
\begin{tabular}{|c|c|c|c|}
	\hline
	$l_r$ & $m / \sqrt{\sigma}$ &  $(E - 4 m c^2)/E_0$ & $\Gamma$ / $E_0$ \\
	\hline
	0 & 1.0 & 5.001 & 0.039 \\
	 & 1.5 & 5.096 & 0.022 \\
	1 & 1.0 & 5.659 & 0.137 \\
	 & 2.0 & 5.903 & 0.093 \\
	 & 3.0 & 5.990 & 0.075 \\
	 & 4.0 & 6.031 & 0.064 \\
	 & 5.0 & 6.053 & 0.053 \\
	2 & 1.0 & 6.194 & 0.586 \\
	 & 2.0 & 6.510 & 0.259 \\
	 & 3.0 & 6.634 & 0.209 \\
	 & 5.0 & 6.736 & 0.171 \\
	 & 7.0 & 6.777 & 0.162 \\
	\hline
\end{tabular}
\caption{Decay widths as a function of $l_r$, with reduced mass correction.}
\label{widthtable2}
\end{table}

\section{Conclusion and outlook}

We study tetraquarks with a simplified model, where the number of Jacobi variables is reduced. 
Nevertheless our computations are fully unitary, and we are able to study resonances above 
threshold, including their wavefunctions, the associated phase shifts, and their decay widths.

We conclude that with higher orbital angular momentum it is easier to form resonances, thus
the existence of tetraquarks is favoured when the orbital angular momentum is finite.
This is consistent with Karliner and Lipkin 
\cite{Karliner:2003dt}
who proposed an angular momentum mechanism for the binding of pentaquarks. 
Adapting their mechanism to tetraquarks, if we regard the tetraquark as a diquark-diantiquark system,
the angular momentum generates a centrifugal barrier between the diquarks, impeding the
recombination of the quarks with the antiquarks to form mesons, and thereby increasing the stability of the
system. Moreover we find that the angular momentum of the confined coordinate $r$ anticommutes
with the angular momentum of the continuum coordinate $\rho$, and this is crucial to prevent the
fast decay to the continuum.

When a relativistic correction is included in our model, we also find that the formation of resonances is 
favoured by low quark masses. This happens since lower quark masses produce a higher binding energy, 
with a higher meson to quark mass ratio favouring the localization the of resonances or boundstates.
We indeed  observe, for sufficiently large angular momentum ($l_r = 3$) and sufficiently
small quark masses, that we have not only the appearance of resonances but also of bound states.

Extrapolating the results obtained with this simplified potential to the similar 
triple flip-flop potential, and utilizing the typical scales of hadronic physics,
we find plausible the existence of resonances in which the tetraquark component 
originated by a flip-flop potential is the dominant one. The decay of a tetraquark
to a meson pair turns out to lead to a narrow width, of the order of 10 MeV, 
when compared to the typical decay width of hadrons or the order of 100 MeV.
 
For instance in the case of a S=2 and J=2 light tetraquark decaying
to a pair of $\rho$ mesons, the decay width computed here  
is sufficiently small  to be neglected when compared with the effect of the rho
meson decay width on the total decay width of the tetraquark.
Thus this suggests that light tetraquarks exist, with
decay widths comparable to the decay widths of normal
mesonic resonances 

We also remark that the unitarized methods developed here are amenable 
to future studies of the full tetraquark problem with the triple flip-flop potential.


\acknowledgments
We thank George Rupp for useful discussions. We acknowledge the 
financial support of the FCT grants CFTP, CERN/FP/109327/2009 and
CERN/FP/109307/2009.


\end{document}